\documentclass[letterpaper, 10 pt, conference]{ieeeconf}

\IEEEoverridecommandlockouts
\usepackage{cite}
\usepackage{comment}
\usepackage[hidelinks]{hyperref}

\usepackage{url}
\usepackage{amsmath, amssymb, amsfonts, bm, dsfont}
\usepackage{stmaryrd} 
\usepackage{algorithmic}
\usepackage{graphicx}
\usepackage{textcomp}
\usepackage{soul}
\usepackage{xcolor}
\usepackage{subcaption}
\def\BibTeX{{\rm B\kern-.05em{\sc i\kern-.025em b}\kern-.08em
    T\kern-.1667em\lower.7ex\hbox{E}\kern-.125emX}}

\newtheorem{remark}{Remark}
    
\begin{document}


\newcommand{\Lagr}{\mathcal{L}}
\newcommand{\one}{\mathds{1}_{m_\text{tot}}}

\newcommand*\TRANS{{\mathpalette\doTRANS\empty}}
\makeatletter
\newcommand*\doTRANS[2]{\raisebox{\depth}{$\m@th#1\intercal$}}
\makeatother

\newcounter{assumption}
\renewcommand{\theassumption}{\arabic{assumption}}

\newcommand{\assumption}[2][]{%
    \refstepcounter{assumption}%
    \par\vspace{0.5em}
    \par\noindent\textbf{Assumption \theassumption:}~#2 
    \par\vspace{0.5em}%
    \ifx#1\empty\else\label{#1}\fi
}

\newcounter{proposition}
\renewcommand{\theproposition}{\arabic{proposition}}

\newcommand{\proposition}[2][]{%
    \refstepcounter{proposition}%
    \par\vspace{0.5em}
    \par\noindent\textbf{Proposition \theproposition:}~#2\hfill$\square$\par\vspace{0.5em}%
    \ifx#1\empty\else\label{#1}\fi
}

\newcounter{problem}
\renewcommand{\theproblem}{\arabic{problem}}

\newcommand{\problem}[2][]{%
    \refstepcounter{problem}%
    \par\vspace{0.5em}
    \par\noindent\textbf{Problem \theproblem:}~#2\hfill$\square$\par\vspace{0.5em}%
    \ifx#1\empty\else\label{#1}\fi
}

\newcounter{theorem}
\renewcommand{\thetheorem}{\arabic{theorem}}

\newcommand{\theorem}[2][]{%
    \refstepcounter{theorem}%
    \par\vspace{0.5em}
    \par\noindent\textbf{Theorem \thetheorem:}~#2\hfill$\square$\par\vspace{0.5em}%
    \ifx#1\empty\else\label{#1}\fi
}


\title{Discrete-time linear quadratic stochastic control with equality-constrained inputs: Application to energy demand response
\thanks{*This work is supported in part by NSERC grant RGPIN-2024-06612 and FRQNT grant 359699 (https://doi.org/10.69777/359699).} 
\thanks{Léo Seugnet and Shuang Gao are with the Department of Electrical Engineering, Polytechnique Montréal, and Groupe d'études et de recherche en analyse des décisions (GERAD), Montréal, QC, Canada. Emails: 
    \texttt{$\{$leo.seugnet,shuang.gao$\}$@polymtl.ca}   
    }
}

\author{Léo Seugnet and Shuang Gao}


\maketitle

\begin{abstract}
We investigate the discrete-time stochastic linear quadratic  control problem for a population of cooperative agents under the hard equality constraint on total control inputs,  motivated by demand response in renewable energy systems.
We establish the optimal solution that respects hard equality constraints for systems with additive noise in the dynamics. The optimal control law is derived using dynamic programming and Karush-Kuhn-Tucker (KKT) conditions, and the resulting control solution depends on a discrete-time Riccati-like recursive equation. 
Application examples of coordinating the charging of a network of residential batteries to absorb excess solar power generation are demonstrated, and the proposed control is shown to achieve exact power tracking while considering individual State-of-Charge (SoC) objectives.
\end{abstract}


\section{Introduction}



With the integration of renewable energy sources such as solar and wind power, modern energy grids are facing growing challenges due to the intermittent nature of these sources \cite{2023Kataray,2024Pallage}. The presence of energy storage units if used properly can help reduce uncertainties as well as accommodate temporary loss of generation capacity on the grids \cite{2023Kataray}. The scheduling of energy usage from the users' side (which is known as demand response) can also help reduce the fluctuations and peak demands (see e.g. \cite{2024Lesage-Landry}).   
The coordination of a group of cooperative energy users can be cast as a stochastic control problem where each user has a local dynamics representing the energy storage level and all users share a global objective of satisfying the demand response request to balance the load and supply. As an approximate model for  demand response problems, we consider a group of users with linear stochastic dynamics and quadratic expected cost in discrete time to establish analytical solutions. 

To reduce uncertainties in the mismatch between the supply and actual demand by collaboratively adjusting the schedules of electricity usages of a large group of users, a key aspect of our formulation is to introduce a hard equality constraint; more specifically, the total charging power of the group of collaborative users matches exactly a prescribed demand profile imposed by the virtual operator.


For {deterministic} linear quadratic (LQ) control problems in discrete time, inequality constraints have been treated in works such as \cite{1998Scokaert,1998Chmielewski,2007Mare,2013Chang}, whereas equality constraints have been addressed in \cite{2007Ko,2010Sideris,2019Laine,2025Laurenzi}. 
Constrained LQ problems can be solved using Quadratic Programming (see e.g. \cite{1998Scokaert,2019Laine}), however the computational complexity of directly applying Quadratic Programming to LQ problems scales cubically with respect to the time horizon (or the length of the trajectories) as discussed in \cite{2019Laine, 2010Sideris}.   
More efficient alternatives based on dynamic programming have been established to solve the LQ control problems subject to inequality constraints (e.g. \cite{1998Chmielewski, 2007Mare, 2020Mitze}) and those subject to equality constraints (see \cite{2007Ko,2010Sideris,2019Laine}).

Stochastic control with constraints in continuous time or discrete time  has been studied by many researchers (see e.g. \cite{1998Chmielewski,2007Ko,2008Krokavec,2011Krokavec,2014Krokavec,2015Hassan,2016Chen,2020Wu}). In these previous works, the constraints are not the hard equality constraints on control inputs. For instance, when equality constraints are considered for discrete-time stochastic systems, they are imposed on the (expected) state \cite{2008Krokavec,2011Krokavec,2014Krokavec,2007Ko} or the terminal state \cite{2016Chen}. The works \cite{2015Hassan,2020Wu} focused on inequality constraints, which impose upper and lower bounds on the control inputs. 
%

\textbf{Contribution}:
%
This paper addresses the hard equality constraints on the sum of control inputs for stochastic linear quadratic control problems in discrete time. It establishes a  Riccati-like difference equation for the optimal stochastic control solution satisfying the equality constraint, derived using dynamic programming together with the KKT conditions at each dynamic programming step. 
Such a solution method is extended to address intermittent equality constraints on the sum of the control inputs, where the hard equality constraint is active only during pre-specified time intervals. 
Additionally, to ensure smooth control transitions, a ``switched" control strategy is established by solving the associated problems with both soft and hard constraints over different time intervals.



\textbf{Notation}: 
%
Let $\mathbb{R}$ and $\mathbb{N}$ denote the set of real and that of natural numbers, respectively. Let $\mathbb{N}^* = \mathbb{N} \setminus \{0\}$ denote the set of nonzero natural numbers. For any $a, b \in \mathbb{N}$ with $a \le b$, let $\llbracket a, b \rrbracket : =\{a, a+1, \dots, b\}$ denote the discrete interval. For a matrix $A$, $A^\TRANS$ denotes its transpose. For a symmetric matrix $Q$ and a vector $z$, let $|z|^2_{Q} := z^\TRANS Q z$. For any $n\in \mathbb{N}^*$, let $\mathds{1}_n \in \mathbb{R}^{n}$ denote the column vector of all ones and $I_n \in \mathbb{R}^{n\times n}$ the identity matrix. Let $[N]:=\{1,2,\cdots, N\}$. Lastly, 
$\text{diag}(M_1, \ldots, M_N)$  denotes the  matrix with diagonal blocks $M_1, \cdots, M_N$  and zero elsewhere. 
 
\section{System model and problem formulation}
\label{Section_2}

\subsection{System model}

Consider $N \in \mathbb{N^*}$ collaborative agents with discrete time dynamics over a finite horizon $\llbracket  0, T \rrbracket $. For an agent $i$, let $x_{i,t} \in \mathbb{R}^{d_x^i}$ (resp. $u_{i,t}\in \mathbb{R}^{d_u^i}$) denote its state (resp. control) with  $d_x^i$ (resp. $d_u^i$) as the dimension. At time $t=0$, the system starts from an initial state $x_{i,0}$ and for $t \in \llbracket 0 ,  T-1 \rrbracket$, the state of agent $i$ evolves according to the  linear dynamics
\begin{equation}\label{agent_dynamics}
    x_{i,t+1} = A_i x_{i,t} + B_i u_{i,t} + w_{i,t},
\end{equation}
where $A_i$ and $B_i$ are  matrices of appropriate dimensions and $w_{i,t} \in \mathbb{R}^{d_x^i}$ is the process noise at time $t$.

Let $x_t = [x_{1,t} , \cdots, x_{N,t}]^\TRANS $, $u_t = [u_{1,t} , \cdots, u_{N,t}]^\TRANS $ and $w_t = [w_{1,t} , \cdots, w_{N,t}]^\TRANS$. The compact representation of the $N$-agent system is then given by
\begin{equation}\label{global_dynamics}
    x_{t+1} = A x_t + B u_t + w_t,
\end{equation}
where $A = \text{diag}(A_1, \dots, A_N)$,  $B = \text{diag}(B_1, \dots, B_N)$,
and $A_i$ and $B_i$ are system parameters of  agent $i$ in (\ref{agent_dynamics}). We note $n_\text{tot}:=~\sum_{i=1}^{N} d_x^i$ and $m_\text{tot}:=~\sum_{i=1}^{N} d_u^i$ the dimensions of the $N$-agent state $x_t$ and control input $u_t$, respectively.


\assumption[ass:Gaussian]{
 $\{w_t\}_{t\ge 0}$ is an i.i.d. noise sequence with mean zero and 
 finite covariance matrix $W \in \mathbb{R}^{n_\text{tot}\times n_\text{tot}}$.
}

\subsection{System performance and control objective}

At time $t \in \llbracket 0 , T-1 \rrbracket$,  agent $i$ incurs an instantaneous cost 
\begin{equation}
    \ell_i(x_{i,t},u_{i,t}) = |x_{i,t} - r_{i,t}|^2_{Q_i} + |u_{i,t}|^2_{R_i} ,
\end{equation}
and at the terminal time $T$, a terminal cost
\begin{equation}
    \ell_{i,T}(x_{i,T},u_{i,T}) =  |x_{i,T} - r_{i,T}|^2_{Q_{i,T}} ,
\end{equation}
where $Q_i$, $Q_{i,T}$, and $R_i$ are matrices of appropriate dimensions, and $r_{i,t}$ are a given deterministic reference trajectory.

The global system (\ref{global_dynamics}) incurs an instantaneous cost
\begin{equation}
    \ell (x_{t},u_{t}) = \sum_{i=1}^N \ell_i(x_{i,t},u_{i,t}) ,
\end{equation}
and at the terminal time $T$, a terminal cost
\begin{equation}
    \ell_{T}(x_{T}) = \sum_{i=1}^N \ell_{i,T}(x_{i,T}) .
\end{equation}
Let $Q = \text{diag}(Q_1, \dots, Q_N)$, $Q_{T} = \text{diag}(Q_{1,T}, \dots, Q_{N,T})$, and $R = \text{diag}(R_1, \dots, R_N)$.

\assumption[ass:QRsym]{The matrices $Q_i$ and $Q_{i,T}$ are symmetric and positive semi-definite and $R_i$ is symmetric and positive definite.}

Assumptions \ref{ass:Gaussian} and \ref{ass:QRsym} follow standard assumptions in stochastic linear quadratic control problems.

\problem[prob1]{
\textit{Choose a control trajectory $u:\llbracket 0 , T-1 \rrbracket \rightarrow \mathbb{R}^{m_\text{tot}} $ to minimize}
\begin{equation}
    J(u_t) = \mathbb{E} \sum_{t=0}^{T-1} \ell (x_{t},u_{t}) + \mathbb{E} \ell_{T}(x_{T})
\end{equation}
\textit{subject to the dynamics \eqref{global_dynamics} and the equality constraint}
\begin{equation}\label{constraint}
    \one^\TRANS u_t = c_t, \quad  \quad  \forall t \in \llbracket 0 , T-1 \rrbracket
\end{equation} 
\textit{where $c_t \in \mathbb{R} $ represents the total consumption requirement at time $  t \in \llbracket 0 , T-1 \rrbracket$.}
}

\section{Main results on optimal control solutions}
\label{Section_3}

The optimal solution to Problem \ref{prob1} is obtained using dynamic programming together with the KKT conditions at each dynamic programming step, and is given as follows.

\theorem[theorem1]{
\textit{Let Assumptions \ref{ass:Gaussian} and \ref{ass:QRsym} hold.  Let $P : \llbracket 0 , T \rrbracket \rightarrow \mathbb{R}^{n_\text{tot} \times n_\text{tot}} $ and $s  : \llbracket 0 , T \rrbracket \rightarrow \mathbb{R}^{n_\text{tot}}$ be the solutions of the following backward recursions}
\begin{align}
    P_{t} &= Q +  A^\TRANS P_{t+1} A  -  A^\TRANS P_{t+1} B \Gamma_t B^\TRANS P_{t+1} A, \label{eq:GRiccati} \\
    s_{t} &= \left[ A^\TRANS - A^\TRANS P_{t+1} B \Gamma_t B^\TRANS  \right] s_{t+1} +  A^\TRANS P_{t+1} B \gamma_t - Q r_{t}, \label{eq:s}
\end{align}
\textit{with the final conditions $P_{T}=Q_T$ and $s_T = - Q_T r_{T}$}, 
\textit{where}
\begin{align*}
    \Gamma_t &= \Omega_{t}^{-1} -  \frac{\Omega_{t}^{-1} \one \one^\TRANS \Omega_{t}^{-1}}{\one^\TRANS \Omega_{t}^{-1} \one},  \quad 
    \gamma_t = \frac{\Omega_{t}^{-1} \one c_t}{\one^\TRANS \Omega_{t}^{-1} \one}, \\
    \Omega_{t} &= R + B^\TRANS P_{t+1} B.
\end{align*}
\textit{Then the optimal control strategy for Problem \ref{prob1} is given by}
\begin{equation}\label{optimal_control}
    u_t = - \Gamma_t ( B^\TRANS P_{t+1} A x_t + B^\TRANS s_{t+1}) + \gamma_t
\end{equation}
\textit{for all $t  \in \llbracket 0 , T-1 \rrbracket$.}
}

\noindent
\textsc{PROOF: }{The cost of Problem \ref{prob1} can be written as follows 
\begin{equation*}
    J(u_t) =  \mathbb{E} \sum_{t=0}^{T-1} \left[ |x_t - r_t|^2_Q + |u_t|^2_R \right]  + \mathbb{E} |x_T - r_T|^2_{Q_T}
\end{equation*}
Let $V_{t}(z)$ be the optimal cost-to-go (or the value function) at $t$ starting from $x_t=z$, defined as
\begin{equation*}
    V_t(z) = \min_{u_t, \dots, u_{T-1}} \mathbb{E} \left[\sum_{\tau=t}^{T-1} \ell(x_\tau, u_\tau) + \ell_T(x_T) \Big\vert  x_t = z \right] .
\end{equation*}
Based on the terminal cost, the value function at the terminal time $T$ satisfies $V_T(z) = |z - r_T|^2_{Q_T} = z^\TRANS Q_T z - 2 r_T^\TRANS Q_T z + r_T^\TRANS Q_T r_T$. We proceed by backward induction, assuming the value function at time $t$ is of form 
\begin{equation}\label{value_form}
    V_{t}(z) = z^\TRANS P_{t} z + 2 s_{t}^\TRANS z + q_{t}
\end{equation}
which holds for $t=T$ with $P_T = Q_T$ and $s_T = -Q_T r_T$. The Bellman recursion is as follows
\begin{equation*}
    V_t(z) = \min_{u} \left\{ |z - r_t|^2_Q + |u|^2_R + \mathbb{E} \left[ V_{t+1}(Az + Bu + w_t) \right] \right\}
\end{equation*}
subject to the constraint $\one^\TRANS u_t = c_t$. Assuming the previous form of the value function (\ref{value_form}), the Bellman recursion can be written as
\begin{align*}
    V_t(z) &= |z - r_t|^2_Q + |z|^2_{A^\TRANS P_{t+1} A} + 2 s_{t+1}^\TRANS A z + \text{tr}(WP_{t+1}) \\
    & \qquad+ q_{t+1} + \min_{u} \left\{ |u|^2_{\Omega_t} + 2 u^\TRANS f_{t}(z) \right\}
\end{align*}
using $\mathbb{E}(w_{t+1}^\TRANS P_{t+1} w_{t+1})=\text{tr}(WP_{t+1})$, where $\Omega_{t} := (R + B^\TRANS P_{t+1} B)$ and $f_{t}(z) = \big(  B^\TRANS P_{t+1} A z + B^\TRANS s_{t+1} \big) $.
To respect the constraint $\one^\TRANS u_t = c_t$, we introduce the following Lagrangian function
\begin{equation*}
    \Lagr (u, \lambda) =  u^\TRANS \Omega_{t} u  + 2 u^\TRANS f_{t}(z) + \lambda (\one^\TRANS u - c_t ).
\end{equation*}
The KKT conditions \cite{1951kuhn} give $2 \Omega_t u + 2 f_t(z) + \lambda \one = 0$, which implies
$u = -\Omega_t^{-1} (f_t(z) + \frac{\lambda}{2}\one )$. Substituting this into the constraint $\one^\TRANS u_t = c_t$ yields the Lagrange multiplier term
\begin{equation*}
    \frac{\lambda}{2} = - \frac{\one^\TRANS \Omega_t^{-1} f_t(z) + c_t}{\one^\TRANS \Omega_t^{-1} \one}.
\end{equation*}
Plugging this back into the expression for $u$ yields the optimal action  $u_t^* = - \Gamma_t ( B^\TRANS P_{t+1} A x_t + B^\TRANS s_{t+1}) + \gamma_t$ where
\begin{align*}
    \Gamma_t &= \Omega_{t}^{-1} -  \frac{\Omega_{t}^{-1} \one \one^\TRANS \Omega_{t}^{-1}}{\one^\TRANS \Omega_{t}^{-1} \one},  
    &\gamma_t = \frac{\Omega_{t}^{-1} \one c_t}{\one^\TRANS \Omega_{t}^{-1} \one}.
\end{align*}


Since $R$ is assumed to be positive definite and $P_t$ is positive semidefinite  (see Lemma~1 in the Appendix \ref{Appendix_B}), we obtain $\Omega_t = R + B^\TRANS P_{t+1} B > 0$. Hence that optimization problem at each dynamic programming step is strictly convex and  the control identified using KKT at each time step is indeed the unique optimal control action.

Finally, by substituting the optimal control $u_t^*$ back into the Bellman equation we get the following equality
\begin{align*}
    V_{t}(z) &= z^\TRANS \Big[ Q + A^\TRANS P_{t+1} A  \Big] z + 2 (  s_{t+1}^\TRANS A - r_{t}^\TRANS Q ) z + \theta_{t+1}  \\
    & \quad + \Big[ -\Gamma_t f_{t}(z) + \gamma_t \Big]^\TRANS \Omega_{t} \Big[ -\Gamma_t f_{t}(z) + \gamma_t \Big] \\
    & \quad + 2 \Big[ -\Gamma_t f_{t}(z) + \gamma_t \Big]^\TRANS f_{t}(z) ,
\end{align*}
where $\theta_{t+1} := q_{t+1} +  r_{t}^\TRANS Qr_{t} + \text{tr}(WP_{t+1}) $. By substituting the expression for $f_t(z)$ and regrouping terms based on their dependency on $z$, we arrive at the following form
\begin{align*}
    V_{t}(z ) &= z^\TRANS \Big[ Q + A^\TRANS P_{t+1} A - A^\TRANS P_{t+1} B \Gamma_t B^\TRANS P_{t+1} A \Big] z \\
    & \quad + 2 \Big[ \left( A^\TRANS - A^\TRANS P_{t+1} B \Gamma_t B^\TRANS  \right) s_{t+1} +  A^\TRANS P_{t+1} B \gamma_t \\
    & \quad -  Q r_{t} \Big]^\TRANS z + q_{t+1} + r_{t}^\TRANS Qr_{t} + \text{Tr}(WP_{t+1})  \\
    & \quad + \gamma^\TRANS \Omega_{t} \gamma_t + 2 s_{t+1}^\TRANS B \gamma_t - s_{t+1}^\TRANS B \Gamma_t B^\TRANS s_{t+1}.
\end{align*}
We recognize the form $V_t(z) = z^\TRANS P_t z + 2s_t^\TRANS z + q_t$ which confirms our previous assumption on the form of $V_t$ in \eqref{value_form}. The identification of the terms $P_t$ and $s_t$ gives us the following recursion equations \eqref{eq:GRiccati} and \eqref{eq:s}.
}

\begin{remark}\label{remark1}
    
    The structure of the optimal control law (\ref{optimal_control}) admits an elegant geometric interpretation. The term $-\Omega_{t}^{-1}( B^\TRANS P_{t+1} A x_t + B^\TRANS s_{t+1})$ is the optimal unconstrained control law. The constrained problem requires this control action to be projected onto the affine subspace defined by the hard constraint (\ref{constraint}). The operators $\Gamma_t$ and $\gamma_t$ perform this orthogonal projection in the Hilbert space of control actions 
    endowed with this $\Omega_t^{-1}$-weighted inner product, ensuring the solution lies on the required affine manifold to meet the equality constraint $c_t$. 
    
    The equation \eqref{eq:GRiccati} is similar to but different from the generalized discrete-time Riccati equation \cite{ferrante2013generalised,ionescu1996generalized}. 
\end{remark}

\begin{remark}
    Although the problem is formulated with matrices $A$, $B$, $W$, $Q$, $Q_T$ and $R$ being block-diagonal, all the proof steps and hence the results in Theorem~\ref{theorem1} apply to problems with general cases with non-block-diagonal matrices. 
\end{remark}

\begin{remark}
    It is important to note that the feasibility of the constraint (\ref{constraint}) is physically limited by the aggregate capacity of the $N$ participating agents. For example,  if  agent~$i \in [N]$ has minimum and maximum (scalar) control actions denoted by $u_{i,\min}$ and $u_{i,\max}$ respectively, then the constraint $c_t$ must lie within the aggregate operational range of the system, i.e., $\sum_{i=1}^{N} u_{i,\min} \le c_t \le \sum_{i=1}^{N} u_{i,\max}$. Consequently, a larger number of users $N$ may provide greater flexibility and  enable the system to satisfy a wider range of constraints $c_t$.
\end{remark}

\begin{remark}
Considering the case  with $d_x^i=n$ and $d_u^i = m$ for all $i \in [N]$.
    The computation of the control in Theorem~\ref{theorem1} involves a computation complexity $\mathcal{O}\left( T N^3 \left( n^3 + m^3 + n^2m + nm^2 \right) \right)$ floating point operations (flops). 
     The  complexity scales linearly with the time horizon $T$. In contrast, applying generic Quadratic Programming 
     to solve constrained LQ control problems incurs a computation complexity $\mathcal{O}(T^3N^3(2n+m)^3)$ flops scaling cubically with respect to the time horizon $T$ as discussed  in \cite[p.~553]{boyd2004convex} and \cite{2019Laine, 2010Sideris}.
    In demand response problems, such a complexity reduction  with respect to the time horizon allows a longer planning horizon.
\end{remark}


\section{Generalizations: Intermittent Constraints}
\label{Section_4}

The framework developed for hard equality constraints can be adapted to address  scenarios with intermittent constraints. In this section we explore two such generalizations. First, we consider the case where the hard equality constraint is only active during specific, predetermined time intervals. Second, we consider the case that  includes intermittent soft constraints where deviations from the target are penalized rather than strictly forbidden, along with intermittent hard equality constraints.

\subsection{Intermittent hard constraint}\label{refS}
Denote the set of time steps with active  constraint by
$$ \mathbb{S} := \{t \in \llbracket 0, T-1\rrbracket \mid \sigma(t) = 1\} $$
where $\sigma: \llbracket 0, T-1\rrbracket \to \{0, 1\}$ is a binary-valued signal satisfying
\begin{equation} \label{eq:sigma}
\sigma(t) =
    \begin{cases}
       1 & \text{if the hard constraint is active at time $t$}; \\
       0  & \text{otherwise}.
    \end{cases}
\end{equation}

\problem[prob2]{
\textit{Choose a control trajectory $u:\llbracket 0 , T-1 \rrbracket \rightarrow \mathbb{R}^{m_\text{tot}} $ to minimize
\begin{equation}
    J(u_t) = \mathbb{E} \sum_{t=0}^{T-1} \ell (x_{t},u_{t}) + \mathbb{E} \ell_{T}(x_{T})
\end{equation} subject to the dynamics \eqref{global_dynamics} and the intermittent equality constraint
\begin{equation}\label{constraint2}
    \one^\TRANS u_t = c_t, \quad  \forall t \in \mathbb{S} \subseteq \llbracket 0 , T -1\rrbracket
\end{equation}
\textit{where $c_t \in \mathbb{R} $ represents the intermittent total consumption requirement at time $  t \in \mathbb{S} \subseteq \llbracket 0 , T-1 \rrbracket$.}
}}

The solution to Problem~\ref{prob2} builds  upon the solution  in Theorem~\ref{theorem1}.  Since the hard equality constraint is only active for time steps within the set $\mathbb{S}$, the optimal control law adopts a switched structure For any time $t \in \mathbb{S}$, the control is derived using the constrained formulation and is similar to the solution of Problem~\ref{prob1}; for any time $t \notin \mathbb{S}$, the problem reduces to a standard unconstrained stochastic LQ problem. 
Proposition~\ref{theorem2} below formalizes this mechanism by presenting a control law where matrices the $\Gamma_t$ and $\gamma_t$ switch between their constrained and unconstrained forms depending on the activation status of the constraint.  

\proposition[theorem2]{
\textit{Let Assumptions \ref{ass:Gaussian} and \ref{ass:QRsym} hold and let $P: \llbracket 0 , T \rrbracket \rightarrow \mathbb{R}^{n_\text{tot} \times n_\text{tot}} $ and $s  : \llbracket 0 , T \rrbracket \rightarrow \mathbb{R}^{n_\text{tot}}$ be the solutions of the following backward recursions}
\begin{align}
    P_{t} &= Q +  A^\TRANS P_{t+1} A  -  A^\TRANS P_{t+1} B \Gamma_t B^\TRANS P_{t+1} A, \\
    s_{t} &= \left[ A^\TRANS - A^\TRANS P_{t+1} B \Gamma_t B^\TRANS  \right] s_{t+1} +  A^\TRANS P_{t+1} B \gamma_t - Q r_{t},
\end{align}
\textit{with the final condition $P_{T}=Q_T$ and $s_T = - Q_T r_{T}$} 
\textit{where}
\begin{align*}
    \Gamma_t &=
    \begin{cases}
        \Omega_{t}^{-1} - \dfrac{\Omega_{t}^{-1} \one \one^\TRANS \Omega_{t}^{-1}}{\one^\TRANS \Omega_{t}^{-1} \one}, & \text{ if } \sigma(t) = 1 \\[1.2em]
        \Omega_{t}^{-1}, & \text{ if } \sigma(t) = 0
    \end{cases} \\[1em]
    \gamma_t &=
    \begin{cases}
        \dfrac{\Omega_{t}^{-1} \one c_t}{\one^\TRANS \Omega_{t}^{-1} \one}, & \text{ if } \sigma(t) = 1 \\[1.2em]
        0_{m_\text{tot} \times 1}, & \text{ if } \sigma(t) = 0
    \end{cases} \\[1em]
    \Omega_{t} &= R + B^\TRANS P_{t+1} B. 
\end{align*}
\textit{Then the stochastic optimal control strategy for Problem \ref{prob2} is given by}
\begin{equation}\label{optimal_control_2}
    u_t = - \Gamma_t ( B^\TRANS P_{t+1} A x_t + B^\TRANS s_{t+1}) + \gamma_t,
\end{equation}
for all $t  \in \llbracket 0 , T-1 \rrbracket$. 
}

\noindent
\textsc{PROOF: }{
The optimal control law is derived by considering two cases based on the activity of the constraint.

For any time $t \in \mathbb{S}$, the equality constraint is active. 
The derivation of the optimal control law and the corresponding recursions for $P_t$ and $s_t$ follows directly from the proof steps of Theorem~\ref{theorem1}.
For any time $t \notin \mathbb{S}$, the constraint is inactive. The problem reduces to a standard unconstrained stochastic LQ problem. The optimal control law for these time steps is given by $u_t^* = -\Omega_t^{-1}(B^T P_{t+1} A x_t + B^T s_{t+1})$, which corresponds to setting $\Gamma_t = \Omega_t^{-1}$ and $\gamma_t = 0$. Substituting this unconstrained solution into the Bellman equation yields the standard Riccati recursions for $P_t$ and $s_t$ as stated in the proposition for $t \notin \mathbb{S}$.

The minimization at each dynamic programming step is a convex optimization problem with or without linear constraints, and hence the KKT conditions applied are  sufficient for optimality.
Combining these two cases above gives the desired forms for $\Gamma_t$ and $\gamma_t$, and completes the proof.
 ~ \hfill $\blacksquare$
}
\subsection{Intermittent soft and hard constraints}
Building on the intermittent control strategy presented previously, this section introduces a modified approach. Instead of having no constraint when the hard equality constraint is inactive, a soft constraint can be applied in the form of a quadratic penalty term in the cost function. Such formulation does not strictly guarantee the constraint is met at all times but penalizes deviations from the target,  leading to smoother control actions.

More specifically, at time $t \in \llbracket 0, T-1 \rrbracket$, the global system (\ref{global_dynamics}) incurs an instantaneous cost
\begin{equation}\label{cost_prob3}
    \ell^{s}(x_{t},u_{t}) := \ell(x_{t},u_{t}) + \eta(1-\sigma(t))(\one^\TRANS u_t - c_t)^2
\end{equation}
with $\sigma(t)$ defined in \eqref{eq:sigma},
and at the terminal time $T$, the system incurs a terminal cost
\begin{equation}
    \ell_{T}^{s}(x_{T}) = \ell_{T}(x_{T})
\end{equation}
where $\eta \ge 0$ is a scalar penalty weight.

\problem[prob3]{
\textit{Choose a control trajectory $u:\llbracket 0 , T-1 \rrbracket \rightarrow \mathbb{R}^{m_\text{tot}} $ to minimize}
\begin{equation}
    J(u_t) = \mathbb{E}\sum_{t=0}^{T-1} \ell^{s} (x_{t},u_{t}) + \mathbb{E}\ell_{T}^{s}(x_{T})
\end{equation}
\textit{while respecting the following constraint}
\begin{equation}\label{constraint3}
    \one^\TRANS u_t = c_t, \quad  \forall t \in \mathbb{S} \subseteq \llbracket 0 , T-1 \rrbracket
\end{equation}
\textit{where $c_t \in \mathbb{R} $ represents the intermittent total consumption requirement at time $  t \in \mathbb{S} \subseteq \llbracket 0 , T-1 \rrbracket$.}
}

Compared to the previous problems, Problem~\ref{prob3} incorporates an additional quadratic penalty term $\eta (1-\sigma(t)) (\one^\TRANS u_t - c_t)^2$ directly into the cost function to reduce the mismatch between the sum of the control and the required consumption.  
Proposition~\ref{theorem3} below provides the resulting stochastic optimal control solutions. 

\proposition[theorem3]{
\textit{Let Assumptions \ref{ass:Gaussian} and \ref{ass:QRsym} hold.  Let  $P : \llbracket 0 , T \rrbracket \rightarrow \mathbb{R}^{n_\text{tot} \times n_\text{tot}} $ and $s  : \llbracket 0 , T \rrbracket \rightarrow \mathbb{R}^{n_\text{tot}}$ be the solutions of the following backward recursions}
\begin{align}
    P_{t} &= Q +  A^\TRANS P_{t+1} A  -  A^\TRANS P_{t+1} B \Gamma_t B^\TRANS P_{t+1} A \\
    s_{t} &= \left[ A^\TRANS - A^\TRANS P_{t+1} B \Gamma_t B^\TRANS  \right] s_{t+1} +  A^\TRANS P_{t+1} B \gamma_t - Q r_{t}
\end{align}
\textit{with the final condition $P_{T}=Q_T$ and $s_T = - Q_T r_{T}$} 
\textit{where}
\begin{align*}
    \Gamma_t &=
    \begin{cases}
        \Omega_{t}^{-1} - \dfrac{\Omega_{t}^{-1} \one \one^\TRANS \Omega_{t}^{-1}}{\one^\TRANS \Omega_{t}^{-1} \one}, & \text{ if } \sigma(t) = 1 \\[1.2em]
        \Pi_{t}^{-1}, & \text{ if } \sigma(t) = 0
    \end{cases} \\[1em]
    \gamma_t &=
    \begin{cases}
        \dfrac{\Omega_{t}^{-1} \one c_t}{\one^\TRANS \Omega_{t}^{-1} \one}, & \text{ if } \sigma(t) = 1 \\[1.2em]
        \eta c_t \Pi_t^{-1} \one, & \text{ if } \sigma(t) = 0
    \end{cases} \\[1em]
    \Omega_{t} &= R + B^\TRANS P_{t+1} B, \quad 
    \Pi_{t} = R + B^\TRANS P_{t+1} B + \eta \one \one^\TRANS.
\end{align*}
\textit{Then the optimal control strategy for Problem \ref{prob3} is given by}
\begin{equation}\label{optimal_control_3}
    u_t = - \Gamma_t ( B^\TRANS P_{t+1} A x_t + B^\TRANS s_{t+1}) + \gamma_t
\end{equation}
for all $t  \in \llbracket 0 , T-1 \rrbracket$. 
}

For the detailed proof, please refer to the Appendix \ref{Appendix_A}.

\section{Numerical examples for demand response}
\label{Section_5}

Consider an illustrative example where the   energy generated by a solar farm is about to reach its local storage  capacity and  excess solar power need be distributed to users in an energy network. We apply our control strategies to allocate solar power to a network of $N=50$ residential batteries where each battery unit has a  capacity limit of 80 kWh. The batteries are equally divided into two classes ($\alpha$ and $\beta$) with distinct SoC targets of 80\% and 40\% to model diverse household needs. 
The state $x_{i,t} \in \mathbb{R}$ (with $ d_x^i=1$) and control input $u_{i,t}\in \mathbb{R}$ (with $d_u^i=1)$ represent respectively the battery level in kWh and power consumption in kW of battery unit $i$. The power generation data is taken from the Canadian Weather Energy and Engineering Datasets (CWEEDS) from the climate station located in Montreal-East \cite{CWEEDS}. We consider a solar power plan with a panel area of 1000 m² and time span of $T=24$ hours (a full day) with a period of $1$ hour. The stochastic noise $\{ w_{i,t} \}_{t\geq 0}$ is  Gaussian process with mean zero and $\mathbb{E}[w_i^2] = 3$ (kWh)$^2$ and is assumed to be independent among agents. 


Consider a demand response call that imposes  the  equality constraint $\one^\TRANS u_t = c_t$, where $c_t$ is the (predicted) excess power of the solar farm (in kW) and $\one^\TRANS u_t$ is the total power consumption of the $N$ residential batteries (in kW). 
The parameters for user $i$ are as follows:   
\begin{align*}
    & A_i \in [0.96, 0.99], \;
    B_i = 1 \text{ kWh/kW}, \;
    \\
    &Q_i = Q_{i,T} = I_{d_x^i}, \;
    R_i = 0.01 I_{d_u^i}.
\end{align*}
We assume that $A_i$ is uniformly distributed between 0.96 and 0.99 and that each user has an initial SoC uniformly distributed between 40\% and 60\% of its 80 kWh capacity. 

\subsection{Non-intermittent hard constraint}

First, we solve Problem~\ref{prob1} using Theorem~\ref{theorem1}, where the constraint~(\ref{constraint}) is active the full duration of the simulation. 

\begin{figure}[!h]
  \begin{subfigure}[c]{.5\linewidth}
    \centering
    \includegraphics[width=\linewidth]{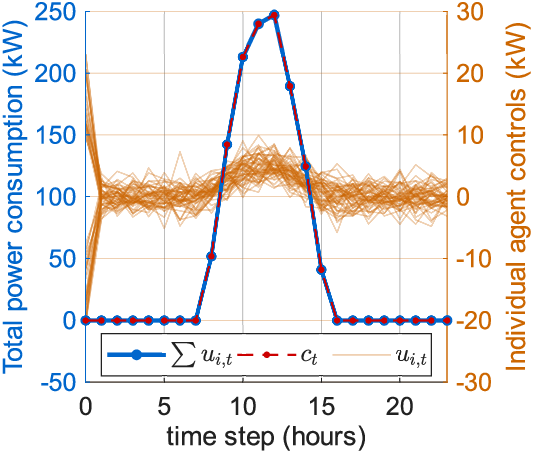}
    \caption{Constraint verification}
    \label{fig:constraint_verification}
  \end{subfigure}
  \hfill
  \begin{subfigure}[c]{.46\linewidth}
    \centering
    \includegraphics[width=\linewidth]{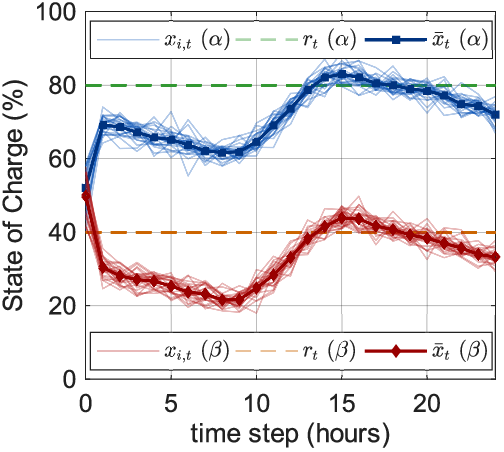}
    \caption{Reference tracking}
    \label{fig:ref_tracking}
  \end{subfigure}
  \caption{Hard non-intermittent constraint}
  \label{fig1}
\end{figure}

The results in Fig.~\ref{fig1}(a) demonstrate that the total power consumed by the network of residential batteries accurately matches the available solar generation at each time step, thereby satisfying the hard equality constraint. Furthermore, Fig.~\ref{fig1}(b) shows that the SoCs of the two classes $\alpha$ and $\beta$, diverge after $t=1$ to successfully track their respective nominal targets. 
\subsection{Intermittent hard constraint}
Consider the case where the constraint~(\ref{constraint2}) is respected intermittently when the excess power is generated by the solar power plant. 
Problem~\ref{prob2} is then solved using Proposition~\ref{theorem2} where the hard constraint is active only for $t\in \mathbb{S} \subseteq \llbracket 0 , T-1 \rrbracket$ where $\mathbb{S} = \{ t \in \llbracket 0 , T-1 \rrbracket \mid c_t \neq 0 \}$. 

\begin{figure}[!h]
  \begin{subfigure}[c]{.5\linewidth}
    \centering
    \includegraphics[width=\linewidth]{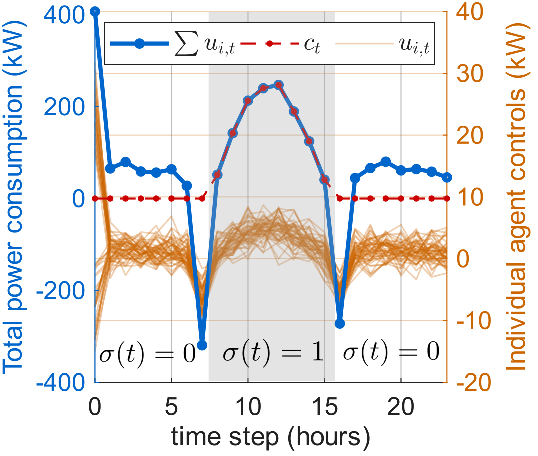}
    \caption{Constraint verification}
    \label{fig:constraint_verification_2}
  \end{subfigure}
  \hfill
  \begin{subfigure}[c]{.46\linewidth}
    \centering
    \includegraphics[width=\linewidth]{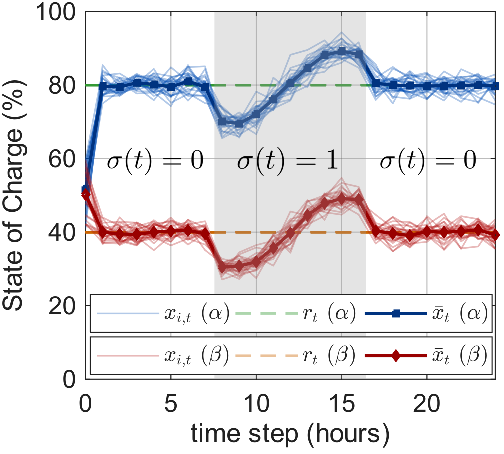}
    \caption{Reference tracking}
    \label{fig:ref_tracking_2}
  \end{subfigure}
  \caption{Hard intermittent constraint}
  \label{fig2}
\end{figure}


The results in Fig.~\ref{fig2} show that when $t \in \mathbb{S}$, the constraint is verified, while for $t \notin \mathbb{S}$$(\sigma (t)=0)$, the individual SoC tracking for classes $\alpha$ and $\beta$ is more accurate. The reason that the individual's tracking performance in Fig.~\ref{fig1} is less satisfactory compared in Fig.~\ref{fig2} in the unconstrained periods is due to the trade-off within the control problem formulation: the equality constraint forces the optimal control inputs to project onto a specific affine subspace (see Remark~\ref{remark1}), removing operational degrees of freedom that would allow the control effort to better track the agents' individual SoC objectives. However, we observe that the transition between the unconstrained and constrained periods results in a sharp negative power peak, as seen in Fig.~\ref{fig2}(a), which implies an undesirable discharge to other local storage devices or flexible loads. This negative peak is reflecting the controller's anticipatory behavior to make ``room" in the batteries so that the impending solar charging doesn't cause the agents’ SoC trajectories to overshoot their objectives.

\subsection{Intermittent soft and hard constraints}

To resolve the issue of the negative power peak (illustrated in Fig.~\ref{fig2}(a)), we use a switched strategy; more specifically, we  activate the hard constraint when $t\in \mathbb{S}$ and switch to a soft constraint when $t \notin \mathbb{S}$, as described in Proposition~\ref{theorem3}. In this formulation, the choice of the penalty weight $\eta$ is critical for balancing the trade-off between enforcing the soft constraint and achieving individual agent tracking objectives. A small value of $\eta$ would place a lower penalty on deviations from the target consumption $c_t$, allowing agents to prioritize tracking their individual SoC objectives more closely, but at the risk of insufficient smoothing during transition periods as seen in Fig.~\ref{fig2}. Conversely, a very large $\eta$ would cause the soft constraint to approximate a hard constraint. 

\begin{figure}[!h]
  \begin{subfigure}[c]{.5\linewidth}
    \centering
    \includegraphics[width=\linewidth]{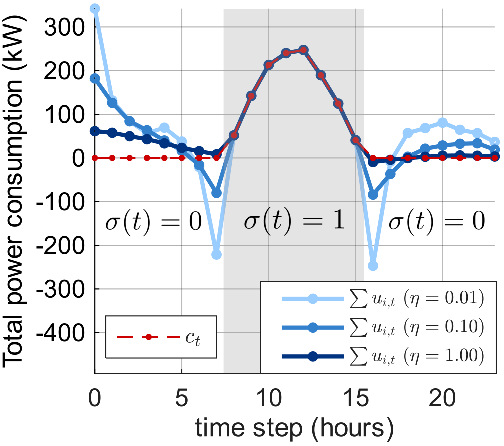}
    \caption{Constraint verification}
    \label{fig:constraint_verification_4}
  \end{subfigure}
  \hfill
  \begin{subfigure}[c]{.46\linewidth}
    \centering
    \includegraphics[width=\linewidth]{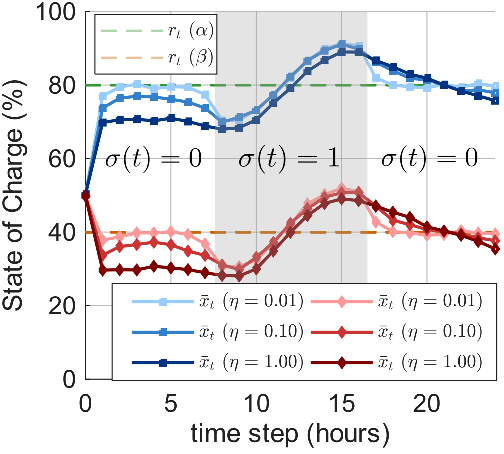}
    \caption{Reference tracking}
    \label{fig:ref_tracking_4}
  \end{subfigure}
  \caption{Soft and hard intermittent constraint}
  \label{fig3}
\end{figure}

As illustrated in Fig.~\ref{fig3}, with the high value of $\eta=1$, the combination of soft and hard intermittent constraints smooths the transition between the constrained and unconstrained periods. This ``switched" approach eliminates the sharp negative power peaks observed in the case with purely intermittent hard constraint. 
Furthermore, a comparison of Fig. \ref{fig3}(b) with Fig.~\ref{fig2}(b) and Fig.~\ref{fig1}(b) reveals that during the soft constrained periods ($\sigma(t)=0$), the agents' SoC trajectories track their reference paths more accurately as $\eta \to 0$. Specifically, decreasing the penalty parameter $\eta$ shifts the optimization trade-off, causing the individual SoC tracking term to dominate the cost function during unconstrained periods.

\section{Conclusion}
\label{Section_Conclusion}

We solved the discrete-time linear quadratic stochastic control problem with hard equality constraints on the summation of control inputs. 
%
The results are generalized to allow for predetermined switches between hard and soft constraints in order to eliminate  sharp power peaks that otherwise may occur with purely intermittent hard constraints.
Numerical examples in the context of energy demand response demonstrated that the total consumptions of the network of batteries exactly matched the excess solar power generation, satisfying the hard constraint, while individual units tracked their distinct SoC objectives. 
%

The collaborative agents face a trade-off between closely tracking their individual SoC objectives and  satisfying the hard equality constraint. Depending on the constraints and individual objectives, a sufficiently large number of participating agents with flexibility in SoC may be needed, which will be investigated in the future. 
Other future work should investigate  the solution to  scenarios with constraints on low-dimensional subspace with network couplings and constraints with individual charging rates,  formulations where agents have local objectives (similar to mean field games), and switching conditions to activate constraints to ensure desired system behavior. 



\section{Acknowledgment}
The authors gratefully acknowledge Roland Malham\'e and Antoine Lesage-Landry for their insightful discussions, as well as the anonymous reviewers for their valuable feedback.
\bibliographystyle{IEEEtran}
\bibliography{main}


\section*{Appendix}
\label{Section_Appendix}

\subsection{Proof of Proposition \ref{theorem3}}
\label{Appendix_A}

The proof proceeds by considering two cases based on the activation of constraints  determined by the signal $\sigma(t)$.

\subsubsection{Case 1: Active hard constraint ($\sigma(t)=1$)} 

For any time $t$ such that $\sigma (t)=1$, the penalty term in the cost function (\ref{cost_prob3}) is null since $(1-\sigma (t))=0$. The problem thus reduces to minimizing the original cost function subject to the hard equality constraint $\one^{\top}u_{t}=c_{t}$. This scenario is identical to Problem \ref{prob1}. Consequently, the derivation of the optimal control law and the corresponding backward recursions for $P_t$ and $s_t$ follow directly from the proof of Theorem~\ref{theorem1}. This yields the expressions for $\Gamma_t$ and $\gamma_t$ as defined in Proposition~\ref{theorem3} for $\sigma (t)=1$.

\subsubsection{Case 2: Active soft constraint ($\sigma(t)=0$)} 

For any time $t$ such that $\sigma(t)=0$, the hard constraint is inactive, and the cost function includes the quadratic penalty term $\eta(\one^{\top}u_t - c_t)^2$. This transforms the problem into an unconstrained stochastic LQ problem with a modified instantaneous cost.

The Bellman recursion at time $t$ starting at state $z$ satisfies
\begin{align*}
    V_t(z) &= \min_{u} \big\{ |z - r_t|_{Q}^{2} + |u|_{R}^{2} + \eta(\one^{\top} u - c_t)^2 \\
    & \qquad + \mathbb{E} \big[ V_{t+1}(Az + Bu + w_t) \big] \big\}
\end{align*}
Assuming the value function at time $t+1$ is of the form $V_{t+1}(z) = z^\top P_{t+1} z + 2 s_{t+1}^\top z + q_{t+1}$, and expanding the expectation, the Bellman recursion becomes 
\begin{align*}
    V_t(z) &= |z - r_t|^2_Q + \text{tr}(WP_{t+1}) + q_{t+1} + |Az|^2_{P_{t+1}} + \eta c_t^2 \\
    & \qquad + 2s_{t+1}^\TRANS Az + \min_{u} \Big\{ |u|^2_{\Pi_t} + 2u^\TRANS g_t(z) \Big\}
\end{align*}
using $\mathbb{E}(w_{t+1}^\TRANS P_{t+1} w_{t+1})=\text{tr}(WP_{t+1})$, where  $\Pi_{t} := R + B^\TRANS P_{t+1} B + \eta\one\one^\TRANS $ and $g_{t}(z) = B^\TRANS P_{t+1} Az + B^\TRANS s_{t+1} - \eta c_t \one $.
The unconstrained minimization problem is $\min_{u} \{ u^\TRANS \Pi_t u + 2u^\TRANS g_t(z) \}$.
Setting the gradient with respect to $u$ to zero yields $2\Pi_t u + 2g_t(z) = 0$, which gives the optimal control law:
\begin{align*}
    u_t^* &= -M_{t}(B^{\top}P_{t+1}Ax_{t}+B^{\top}s_{t+1})+\gamma_{t}
\end{align*}
where $
    \Gamma_t = \Pi_t^{-1}$ and $ 
    \gamma_t = \eta c_t \Pi_t^{-1} \one.
$
The resulting control at each dynamic programming step is optimal  following the standard LQ control theory. 



Substituting the optimal control law back into the Bellman recursion, we can identify the coefficients for the quadratic and linear terms in $z$.
By matching the coefficients with the assumed form $V_t(z) = z^\TRANS P_t z + 2s_t^\TRANS z + q_t$, we obtain the following recursion equations for $P_t$ and $s_t$:
\begin{align*}
    P_{t} &= Q +  A^\TRANS P_{t+1} A  -  A^\TRANS P_{t+1} B \Pi_t^{-1} B^\TRANS P_{t+1} A \\
    s_{t} &= \left[ A^\TRANS - A^\TRANS P_{t+1} B \Pi_t^{-1} B^\TRANS  \right] s_{t+1} - Q r_{t} \\
    & \qquad + A^\TRANS P_{t+1} B \Pi_t^{-1} \one \eta c_t. 
\end{align*}

This confirms our assumption on the quadratic form of the value function in both cases  and completes the proof by induction.
\hfill $\blacksquare$

\subsection{Properties of the Riccati-like equation}
\label{Appendix_B}
\noindent \textbf{Lemma 1:} 
Let $R>0$ and $Q\geq 0$ and $Q_T\geq 0$. 
Then solution $P_t$ to the following recursive equation
\begin{equation}\label{eq:lemma_Riccati}
    P_{t} = Q +  A^\TRANS P_{t+1} A  -  A^\TRANS P_{t+1} B \Gamma_t B^\TRANS P_{t+1} A ,
\end{equation} with $P_{T}=Q_T$ and
\begin{align*}
    \Gamma_t &= \Omega_{t}^{-1} \left( I -  \frac{ \one \one^\TRANS \Omega_{t}^{-1}}{\one^\TRANS \Omega_{t}^{-1} \one} \right),  
   &  \Omega_{t} = R + B^\TRANS P_{t+1} B,
\end{align*}
remains positive semidefinite for all $t\in \llbracket 0, T\rrbracket$. $\hfill \square$
\\ \\
\textsc{PROOF: } This can be proved by backward induction. At \( t = T \), \( P_T = Q_T \geq  0 \) by assumption.
Assume \( P_{t+1} \geq 0 \). Then we need to show that \( P_t \geq 0 \).
To show this, 
we construct an auxiliary linear quadratic regulation problem from state \( x \in \mathbb{R}^{n_\text{tot}} \) at time \( t \), under dynamics
\[
x_{t+1} = A x_t + B u_t
\]
and  input \( u_t \in \mathbb{R}^{m_\text{tot}} \) satisfying \( \one^\TRANS u_t = 0 \). The cost is
\[
J_t(x,u) : = x^\TRANS Q x + u^\TRANS R u + (A x + B u)^\TRANS P_{t+1} (A x + B u).
\]
which  can be represented as
\[
J_t(x,u) = x^\TRANS Q x + x^\TRANS A^\TRANS P_{t+1} A x + 2 u^\TRANS g + u^\TRANS \Omega_t u,
\]
with $g := B^\TRANS P_{t+1} A x.$ 
To solve the problem
$$
\min_{u: \one^\TRANS u = 0} J_t(x,u)
$$
subject to the linear dynamics above, we consider the following Lagrangian 
$
\mathcal{L}(u, \lambda) = u^\TRANS \Omega_t u + 2 g^\TRANS u + \lambda \one^\TRANS u 
$
and identify the first-order optimality conditions 
\begin{align*}
    2 \Omega_t u + 2 g + \lambda \one &= 0, \quad 
    \one^\top u = 0.
\end{align*}
Since $R>0$ and $P_{t+1}\geq 0$,  we obtain that  $\Omega_{t} = R + B^\TRANS P_{t+1} B>0$.
Hence the optimal control action is given by 
$
u^* = - \Omega_t^{-1} \left( g + \frac{\lambda}{2} \one \right).
$
Plugging into constraint
\[
\one^\TRANS u^* = - \one^\TRANS \Omega_t^{-1} g - \frac{\lambda}{2} \one^\TRANS \Omega_t^{-1} \one = 0.
\]
and solving for \( \lambda \) yields
$
\frac{\lambda}{2} = - \frac{ \one^\TRANS \Omega_t^{-1} g }{ \one^\TRANS \Omega_t^{-1} \one }.
$
Thus,
\[
u^* = -\Gamma_t\, g, \quad\text{where}\quad \Gamma_t = \Omega_t^{-1} - \frac{ \Omega_t^{-1} \one \one^\TRANS \Omega_t^{-1} }{ \one^\top \Omega_t^{-1} \one }.
\]
The minimum value of the cost is hence given by
\[
\begin{aligned}
 & x^\TRANS Q x + x^\TRANS A^\TRANS P_{t+1} A x - g^\TRANS \Gamma_t g \\
 & = x^\TRANS \left( Q + A^\TRANS P_{t+1} A - A^\TRANS P_{t+1} B \Gamma_t B^\TRANS P_{t+1} A \right) x.   
\end{aligned}
\]
Therefore, we get the equation (\ref{eq:lemma_Riccati}).
%
%
%
From the cost minimization interpretation, we have
\[
x^\TRANS P_t x = \min_{u:\, \one^\top u = 0} J_t(x,u)
\]
and since all terms in \( J_t(x,u) \) are nonnegative, the minimum must also be nonnegative. That is
\[
x^\TRANS P_t x = \min_{u:\, \one^\top u = 0} J_t(x,u)  \geq 0,  \quad \forall x.
\]
This implies \( P_t \geq 0 \). Thus, by backward induction, \( P_t \geq 0 \) for all \( t \). \hfill $\blacksquare$

\end{document}